
\magnification 1200

\newdimen\pagewidth \newdimen\pageheight \newdimen\ruleht
\hsize=31pc  \vsize=45pc  \maxdepth=2.2pt  \parindent=19pt
\pagewidth=\hsize \pageheight=\vsize \ruleht=.5pt
\abovedisplayskip=10pt minus 2pt                    
\belowdisplayskip=11pt minus 2pt
\abovedisplayshortskip=8pt minus 2pt
\belowdisplayshortskip=9pt minus 2pt

\baselineskip=15pt minus 0.5pt
\lineskip=2pt minus 0.5pt
\lineskiplimit= 2pt

\mathsurround=1pt


\nopagenumbers
\def\testos{\null}
\def\testod{\null}
\headline={\if T\tpage{\gdef\tpage{F}{\hfil}}
    \else{\ifodd\pageno\rightheadline\else\leftheadline\fi}
   \fi}
\gdef\tpage{T}
\def\rightheadline{\hfil{\tensl\testod}\hfil{\tenrm\folio}}
\def\leftheadline{{\tenrm\folio}\hfil{\tensl\testos}\hfil}



\newcount\numref
\global\numref=1
\newwrite\fileref
\immediate\openout\fileref=ref.tmp
\immediate\write\fileref{\parindent 30pt}
\def\citaref#1{${[\the\numref]}$\immediate\write\fileref
      {\par\noexpand\item{{[\the\numref]\enspace}}}\ignorespaces
      \immediate\write\fileref{{#1}}\ignorespaces
      \global\advance\numref by 1\ignorespaces}


\def\reali{\hbox{\rm I\hskip-2pt\bf R}}

\def\naturali{\hbox{\rm I\hskip-5.5pt\bf N}}

\def\eps{\varepsilon}


\def\bsk{\bigskip}
\def\msk{\medskip}
\def\ssk{\smallskip}
\def\ni{\noindent}


 at 14.40 truept

\def\R{{\cal R}}
\def\xn{x_1, \ldots ,  x_n}
\def\xN{x_1, \ldots ,  x_N}
\def\xntn{x_1 (t_1), \ldots ,  x_n(t_n)}

\def\ddxi{{\partial \over \partial x_i}}
\def\dmps{|\psi_E|^2 dx_1 \ldots dx_n}

\hfill   IFUP-TH 35/2001

\bsk\bsk\bsk
\centerline{M. Correggi}
\msk
\centerline{S.I.S.S.A., Trieste, Italy}
\bsk\bsk
\centerline{G. Morchio}
\msk
\centerline{Dipartimento di Fisica and INFN, Pisa, Italy}
\bsk\bsk\bsk
\centerline{\bf Quantum Mechanics and Stochastic Mechanics}
\centerline{\bf for compatible observables at different times}

\bsk\bsk\bsk\bsk\ni
{\bf Abstract}. Bohm Mechanics and Nelson Stochastic Mechanics are
confronted with Quantum Mechanics in presence of non--interacting
subsystems. In both cases, it is shown that correlations at different
times of compatible position observables on stationary states agree
with Quantum Mechanics only in the case of product wave functions. By
appropriate Bell-like inequalities it is shown that no classical
theory, in particular no stochastic process, can reproduce the quantum
mechanical correlations of position variables of non interacting
systems at different times.

\bsk\bsk\bsk\ni

\vfill\eject

\ni
{\bf 1. Introduction}.

\bsk\ni
Bohm Mechanics \citaref{D. Bohm, A suggested interpretation of the quantum
theory in terms
of \lq\lq Hidden\rq\rq\ variables, I and II, {\it Phys. Rev.} {\bf 85}, 166
(1952)}
\citaref{D. Bohm, {\it The undivided universe: an ontological
interpretation of quantum theory}, Routledge and Kegan, London, 1993}
\  and Nelson's Stochastic Mechanics
\citaref{E. Nelson, Derivation of the Schr\"odinger equation from Newtonian
mechanics, {\it Phys. Rev.} {\bf 150}, 1079 (1966)}
\citaref{E. Nelson, Connection between Brownian motion and quantum
mechanics, in {\it Einstein Symposion}, Springer, Lectures Notes in Physics {\bf
100} (1979)}
\citaref{E. Nelson, {\it Quantum fluctuations}, Princeton University Press,
1985} \  have been
extensively studied as classical dynamical systems associating
classical trajectories, respectively
deterministic and stochastic, to the Schr\"odinger equation.

In both theories, particle positions evolve according to equations
which depend on the Schr\"odinger wave function $\psi (x_1, \ldots ,
x_n, t)$ in such a way that the particle density $\rho(x_1, \ldots ,
x_n , t)$ coincides with $ |\psi( {\bf x}, t)|^2 $ at all times if so
does at some (initial) time.

Both theories have been advocated \citaref{S. Goldstein, Quantum 
Theory without Observers (Part Two), Physics Today,
38 (1998)}\ as perfectly
adequate alternative descriptions of Schr\"odinger Quantum Mechanics.
In particular, it is believed that both theories reproduce all
the predictions of non-relativistic Quantum Mechanics of spinless
particles (QM) which can be expressed in terms of observations of particle
positions [6].

As a matter of fact, coincidence of densities at all times amounts to
coincidence of all (probabilistic) predictions for the observation of
position variables at any given time.  It is a fact that {\it for
generic time evolutions}, there are no further quantum mechanical
predictions for position observables, since position variables at
different times do not commute in general, so that a common
probabilistic interpretation is excluded.  Moreover, for spinless
Schr\"odinger particles, there are no additional observables which are
compatible with all position variables at a given time, so that no
additional prediction can be derived in Quantum Mechanics.

It is therefore concluded that {\it if only position observables are
considered, Bohm Mechanics and Nelson Stochastic Mechanics give
exactly the same predictions as Schr\"odinger Quantum Mechanics}.

However, even restricting the attention to position observables of
spinless Schr\"o\-din\-ger particles, the above conclusion is not as
general as it may appear. In fact, position observables of particles
belonging to non--interacting subsystems clearly commute also {\it at
different times}, so that quantum mechanical predictions for their
common values are perfectly defined and can be confronted with the
corresponding predictions of Bohm and Nelson mechanics. Such a
comparison is clearly relevant also in view of the difficulty of
arguing against the measurement of positions of non--interacting
particles at different times, possibly in very distant regions. On the
contrary, it would be hard to assume that exact simultaneity of
measurements can be achieved.

\msk
Difficulties in the physical interpretation of trajectories,
especially in the presence of non interacting subsystems, have been in
fact pointed out and discussed by Nelson
\citaref{E. Nelson,  Field Theory and the Future of Stochastic Mechanics,
in {\it Stochastic Processes in Classical and Quantum Physics},
Springer, Lecture Notes in Physics {\bf 262}, 1986}\
in terms of problems with \lq\lq separability\rq\rq,
i.e. independence of the stochastic equations of a subsystem from
changes in the interactions within different, non interacting,
possibly very distant subsystems.  Blanchard et al. showed in
ref. \citaref{P. Blanchard, S. Golin, M. Serva, Repeated measurements
in stochastic mechanics, {\it Phys. Rev. D} {\bf 34}, 3732 (1986)}\
that discrepancies between QM and Nelson Mechanics disappear if the
effects of measurements on Nelson equations are taken into account,
through a mechanism analogous to a wave function \lq\lq collapse\rq\rq . 
As a result of their analysis, however, the dynamical equations of a
subsystem depend on measurements performed on different, non
interacting, possibly very distant subsystems, and Nelson's criticism
applies.

In the case of Bohm mechanics, to avoid disagreement with QM, Bell
proposed to include the measurement apparata in the system
\citaref{J. S. Bell, {\it Speakable and unspeakable in quantum mechanics},
Cambridge University Press, Cambridge, 1987},
and to consider only position measurements of pointers, {\it all at the same
\lq\lq final\rq\rq\ time}. Clearly, on one side this results in a
totally unrealistic constraint on measurements, on the other it
forbids any interpretation of trajectories, when the main motivation
for Bohm theory was to reformulate and possibly reinterpretate Quantum
Mechanics in terms of them. A discussion of examples of disagreement
between Bohm Mechanics and QM has been given recently in
refs. \citaref{A. Neumeier, Bohmian Mechanics contradicts Quantum
Mechanics, arXiv \goodbreak quant--ph 0001011 (2000)}
\citaref{P. Ghose, Incompatibility of the De Broglie--Bohm Theory with
Quantum Mechanics, arXiv quant--ph 0001024 (2000); On the Incompatibility
between Standard Quantum Mechanics and Conventional De Broglie--Bohm theory,
arXiv quant--ph 0103126 (2001)}.

\msk
The purpose of the present paper is to take seriously Bohm and Nelson
Mechanics as classical probabilistic theories about trajectories and
to clarify the extent of disagreement with the predictions of QM for
compatible position observables of non interacting subsystems.  We
will show in complete generality that both Bohm Mechanics and Nelson
Stochastic Mechanics agree with QM, for a stationary state, if and
only if the state wave function is a product over the subsystems.
Moreover, we will show that such a situation is not special to Bohm
and Nelson theories, proving that {\it no classical probability
theory} can reproduce the QM predictions for compatible position
observables at different times of a large class of systems. This
implies in particular that the proposal of [8] cannot be
interpreted as a conditioning procedure on a classical probability
theory.

\msk
In Sect. 2 the framework of our arguments will be fixed. We will
recall that the existence of classical theories reproducing all
quantum mechanical predictions for position observables at equal times
follows immediately from the spectral theorem and that, more
generally, any collection of probabilistic predictions for {\it
disjoint} sets of observables can be reproduced by classical theories.
We will conclude that position observables of non interacting
particles at different times give the only relevant test for a
trajectory description of Schr\"odinger QM.

\msk
In Sect. 3 we will discuss the predictions of Bohm Mechanics and Nelson
Stochastic Mechanics for the joint distribution of compatible position
observables at different times, for any stationary state.  In both
cases they turn out to be different from the corresponding Quantum
Mechanical predictions, with the only exception of product wave
functions. In the Appendix, Nelson probability distributions will be
shown to be computable in terms of Quantum Mechanical
evolution at imaginary times, with Dirichlet boundary condition on the
nodes of the wave function. An explicit computation will be performed
for a system of two harmonic oscillators.

\msk
In Sect. 4, by constructing appropriate Bell--Clauser--Horn
inequalities, we will show that no classical theory can reproduce
the quantum mechanical predictions for joint distribution of position
observables of non interacting systems at different times, even in
the case of stationary states of very elementary systems.

\msk
From the above results we conclude that no attempt for a classical
description of Schr\"odinger Quantum Mechanics in terms of particle
trajectories can go beyond the reproduction of equal time correlations
and that it is precisely the restriction to disjoint sets of
compatible observables, rather than the use of positions variables,
that allows for a classical description.

\goodbreak
\bsk\bsk\ni
{\bf 2. Comparison of theories and systems of observables}.

\bsk\ni
Comparison between theories clearly depends in general on the class of
experiments, or \lq\lq observables \rq\rq\ which are considered.
Moreover, since in QM not all pairs of observables can be measured
together, the description of sets of observables must include a notion
of {\it joint measurability}. For the sake of clarity, it is
convenient to partially formalize such notions through the following
definitions (see e.g. refs.
\citaref{I. Pitowski, {\it Quantum Probability--Quantum Logic}, Springer.
Lecture Notes in Physics {\bf 321} (1989)}
\citaref{D. W. Cohen, {\it An Introduction to Hilbert Space and Quantum
Logic},
Springer, 1989}). 

\bsk\ni
{\bf Definition 2.1} {\it A set of observables is a set $X$ 
with a relation ${\cal R}$ satisfying reflexivity, $A {\cal R} A$
$\forall A
\in X$ and symmetry, $A {\cal R} B$ $\Rightarrow$ $B {\cal R} A$. If
$A {\cal R} B$, $A$ and $B$ are called compatible}.

\bsk\ni
The elements $A \in X$ are thought to be associated to experimental
devices, providing, in case the experiment is performed, a real number
$x_A(\eta)$ as \lq\lq the result of the measurement of $A$ in the
experiment $\eta$ \rq\rq .  Compatible experiments can be performed
together; given a subset $Y$ of $X$ consisting of compatible
observables, and $A$  in $Y$, the corresponding 
experiments $\eta_Y$ give rise to functions
$f_A: E_Y \mapsto \reali$. 

\bsk\ni
{\bf Definition 2.2} {\it An experimental arrangement is a
subset $Y$ of $X$ consisting of compatible elements, i.e. 
$A {\cal R} B$ $ \forall A,B \in Y$.
For each experimental arrangement $Y$ there is a set $E_Y$ of
experiments and, for all $A \in Y$, a function $f_A : E_Y \mapsto \reali$}.

\bsk\ni
The functions $f_A(\eta_Y)$, $\eta_Y \in E_Y$ contain all the
information about relations between the results of compatible
experiments, usually interpreted as relations between
observables. The same information is contained in the C*
algebras they generate, and in fact it is usually given, e.g. in QM,
in terms of commutative C* algebras of compatible observables. Such
algebras arise automatically if the observables are assumed to form a
(non--commutative) C* algebra, as in the Haag--Kastler approach to QM
\citaref{R. Haag, {\it Local quantum theory: fields, particles, algebras}, 
Springer, 1993}
\citaref{H. Araki, {\it Mathematical theory of quantum fields},
Oxford University Press, 1999}. 
In particular (see below), the above relations can
be taken as the basis for the discussion of the problems of the
interpretation of QM. We introduce therefore the following notion:

\bsk\ni
{\bf Definition 2.3} {\it A system of observables consists of a set
$X$, a compatibility relation ${\cal R}$, a collection of sets $E_Y$
indexed by all experimental arrangements and a collection of functions
$f_A (\eta_Y)$, $A \in Y$}.

\bsk\ni
Given a system of observables, a theory consists in general of a set
of probability assignments for the values of the functions $f_A$, and
can be formalized as follows:

\bsk\ni
{\bf Definition 2.4} {\it A system of predictions for a system of
observables $(X, {\cal R}, \{ E_Y \}, \{ f_A \})$ is an assignment of
probability measures $ d \mu_Y$ on the spaces $E_Y$ so that all the
functions $f_A(\eta_Y)$, $A \in Y$, are measurable}.

\bsk\ni
The predictions for the mean value of the observable $A$,
measured with the experimental arrangement $Y$, is therefore
$$ \int_{E_Y} \, f_A(\eta_Y) \; d \mu_Y (\eta_Y) \
\ \ \ .  $$ 
Given a theory, different systems of observables can be introduced.
In particular, the choice of a subset $X_1$ of a system $X$ of
observables amounts to restricting the interpretation of the theory to
the system of predictions indexed by the experimental arrangements
contained in $X_1$.

\msk
In the ordinary interpretation of QM it is assumed that observables
correspond to selfadjoint operators in a Hilbert space $\cal H$.
Moreover, it is assumed

\ssk\ni
i) {\it two observables are compatible if and only if the corresponding
operators commute},

\ssk\ni
ii) {\it all selfadjoint operators define observables}.

\ssk\ni
Experimental arrangements are therefore indexed by all commutative
C* subalgebras of $\cal B(H)$, the spaces $E_Y$ being given by their spectra
and $f_A$ by the corresponding Gelfand representations. 

\msk
The above notion of system of predictions is close to the generalized
notion of state introduced by Bell
\citaref{J. S. Bell, On the problem of hidden variables in quantum mechanics, 
{\it Rev. Mod. Phys.}, {\bf 38}, 447 (1967)}.  
It is well known that, by Gleason's theorem
\citaref{A. M. Gleason, Measures on the closed subspaces of a Hilbert space,
{\it Journal of Mathematics and Mechanics} {\bf 6}, 885 (1957)}, no
system of predictions on the system of observables defined by i) and
ii) can be reproduced by any classical theory. i.e. by the
identification of the functions $f_A(\eta_Y)$ with a set of
(measurable) functions $F_A(\xi)$ on a {\it common} measure space, if
the dimension of $\cal H$ is greater than two.  It also follows from
the Bell--Kochen--Specker theorem
\citaref{S. Kochen, E. Specker, The problem of hidden variables in quantum
mechanics, {\it Journal of Mathematics and Mechanics} {\bf 17}, 59
(1967)}\
that no classical representation
exists for any system of predictions on the collection of the
commutative subalgebras of the algebra generated by the angular
momentum operators in the representation with $l=1$.  Similarly, Bell
inequalities imply that no classical representation exists for the
system of predictions defined by suitable quantum mechanical states on
the system of observables defined by appropriate spin observables of
two spin $1/2$ particles.

However, as stressed by Bohm and Bell [9], assumption ii) may be too
general, in particular, angular momentum and spin observables may be
in some sense not admissible, and the choice of a physically motivated
subset of observables may lead to a better motivated and possibly
classical interpretation of QM.  This possibility is exploited by
Bohm's and Nelson's trajectory description of Schr\"odinger Quantum
Mechanics. In terms of the above discussion, Bohm's and Nelson's
theories are compared with QM by substituting assumption ii) above
with

\ssk\ni
iiB) {\it only position operators define observables}.

\ssk\ni
In the following we will assume iiB) and consider, for a system of $N$
particles, only observables consisting of measurable,
bounded functions with compact support 
$f(x_i(t))$, $i = 1 \ldots N$, $t\in \reali$.

To discuss the r\^ole of compatibility relations, we first notice that
the predictions of Bohm and Nelson Mechanics are usually compared
to those of QM for position observables {\it at the same time}.  In
the framework discussed above, such a restriction amounts to assuming
as compatibility relation between position observables the fact that
they are measured at the same time.

There are cases in which such a choice is {\it forced by quantum
mechanical compatibility}, i.e. only functions of position at the same
time define commuting operators. This is in fact the case of a free
particle in one space dimension:

\bsk\ni
{\bf Proposition 2.5} {\it Let $f$ and $g$ be bounded measurable
functions with compact support. If, for some
$t \neq s$, $ [f(x + pt/m ) , g(x + ps/m ))]
= 0 $, then $f=0$ or $g=0$}.

\msk\ni
Proof. Let $\tau \equiv (s-t)/m$.  $ [ f(x + p t/m) , g(x + p s/m) ] = 0
$ is equivalent to $ [ f(x) , g(x + p \tau)] = 0 $. The operator
$$g(x + p\tau + a) =
 e^{-iax/\tau} \ g(x + p \tau) \ e^{iax/\tau} \eqno(2.1) $$
then commutes with $f(x)$ for all $a \in \reali$ and therefore $g(x +
p \tau)$ commutes with $f(x-a)$ for all $a$.  If $f$ is not a
constant, the Von Neumann algebra $F$ generated by $f(x-a)$,
$a \in \reali$ contains a
characteristic function $\chi_I(x)$ of a bounded measurable set $I$,
and therefore all $\chi_I (x-a)$.
It is easy to see that, e.g., $\psi(x) = \exp{-x^2}$ is a cyclic
vector for $F$ in $L^2(\reali)$; in fact, for $\phi(x)$ in $L^2(\reali)$,
$$ 0 = (\chi_I(x-a) \, \psi , \, \phi) =
\int dk \; e^{ika} \; \overline{\tilde{\chi_I}} \, \tilde{(\psi  \phi)}   \
\ \ \  \forall a$$
$\tilde{f}$ denoting the Fourier transform of $f$, implies
$$ \overline{\tilde{\chi_I}} \; \tilde{(\psi  \phi)}   = 0 $$
and therefore, by analiticity of $\tilde{\chi_I}$,
$ \tilde{(\psi  \phi)}  = 0 $, $ \psi  \phi  = 0 $ and $\phi = 0$.
It follows that, since $g(x + p\tau)$ commutes with $F$,
it is a multiplication operator, $g(x + p\tau) = h(x)$
and therefore, by eq.(2.1), $g$ is a constant.

\bsk
The above result does not however extend to more than one dimension,
since, e.g., different components of the position of a free particle
commute and are therefore quantum mechanically compatible also at
different times.
More generally, for particle systems composed of non interacting
subsystems, functions of positions of particles belonging to different
subsystems correspond to commuting operators for all times, and
therefore they are compatible observables according to QM.  For such
systems the choice of equal times as the compatibility
relation between position observables has no clear motivation. Common
probability distributions for such positions are predicted by QM and
can be compared with the prediction of Bohm and Nelson
Mechanics.

Moreover, such a comparison is in a sense the only relevant test for
the above theories. In fact, a compatibility relation consisting of
\lq\lq being observed at the same time\rq\rq\ is clearly a transitive
relation, giving rise to the partition of observables into disjoint
classes; the following Proposition recalls that this fact implies by
itself the existence of a common classical description of any set of
probability distributions for all admitted experimental arrangements,
independently from any other assumption on the observables, in
particular from the above assumption iiB).  In other terms, if the
possibility of measuring observables at different times is discarded,
quantum mechanical predictions for any commuting set of observables at
each time a priori admit a classical representation.

\bsk\ni
{\bf Proposition 2.6} {\it Let $X, \R, \{E_Y\}, \{f_A(\eta_Y) \}$ be a
system of observables, with $E_Y$ compact Hausdorff spaces, and $\{
d\mu_Y \}$ a system of predictions for it consisting of Borel measures.  
If $\R$ is transitive, there is a measure space $\Xi, \, d\nu$ and for each $A $
in $X$ a measurable functions $F_A(\xi)$ such that, for all finite
sequences of observables $A^i$
in an experimental arrangement $Y$ and real
intervals $I^i$,
$$ \mu_Y ( \{ \eta_Y : f_{A^i}(\eta_Y) \in I^i \} ) =
   \int_\Xi d\nu \ \prod_{i=1}^n \chi_{I_i} (F_{A^i} (\xi))
   \ \ \  ,  \eqno(2.2) $$
$\chi_{I} $ denoting the characteristic functions of the interval $I$}.

\msk\ni
Proof. Let $\Xi$ be the topological product of the spaces $E_Y$ with
the product measure $d\nu = \prod_Y d \mu_Y$.  Transitivity of $\R$
implies that experimental arrangements are disjoint sets. Each
observable $A$ belongs therefore to a unique experimental arrangement
$Y_A$.  Denoting points of $\Xi$ as $\xi \equiv \{
\xi_Y \}$, the functions $F_A(\xi) \equiv f_A(\xi_{Y_A})$ are
therefore well defined and
$$ \mu_Y (\{ \eta_Y: f_{A^i} (\eta_Y) \in I^i \} ) = \int_{E_Y}
d\mu_Y \;
\prod_{i=1}^n  \chi_{I_i} \, (f_{A^i} (\xi_Y)) = $$
$$ = \int_\Xi d\nu \ \prod_{i=1}^n \chi_{I_i} \, (F_{A^i} (\xi)) \ \ \ .
\eqno(2.3)  $$

\bsk
Since any quantum mechanical state defines, through the spectral
theorem, a system of predictions, in the sense of Def. 2.4, on any
collection of sets of commuting operators, it follows from Proposition
2.6 that any quantum system has a classical description, in terms of
the classical probability space $\Xi$, if the experimental
arrangements admitted by the choice of the compatibility relation are
disjoint.

In particular, if positions observables are assumed to be compatible
only at the same time, all the corresponding quantum mechanical
predictions are represented, by Proposition 2.6, in terms of a
probability measure in the space of trajectories $ \{ x_t \}$.
Clearly, the measure $d\nu$ constructed in Proposition 2.6 is not
unique, and in fact, for quantum mechanical position observables, both
Bohm and Nelson Mechanics provide valid alternatives. The problem of
the abundance of such alternatives has in fact being addressed
\citaref{E. Deotto, G. C. Ghirardi, Bohmian Mechanics Revisited, {\it
Found. Phys.}  {\bf 28}, 1 (1998)}, and clearly no solution with
observational content exists if correlations at different times are
not taken into account.

We conclude that it is very relevant to extend the comparison of
Quantum Mechanics and Stochastic Mechanics to cases in
which positions at different times can be taken as compatible
observables.

\goodbreak
\bsk\bsk\ni
{\bf 3. Bohm's and Nelson's stochastic predictions for compatible positions
at different times}.

\bsk\ni
We will compare the predictions of Quantum Mechanics, Bohm Mechanics
and Nelson Stochastic Mechanics, in the case of stationary states, for
position variables of non interacting subsystem at different times.
The stationary case turns out to be sufficient for a clear comparison
with Quantum Mechanics, since completely different predictions will result
already for stationary states.

\msk
Quantum mechanical predictions can be characterized in very simple
terms for the following class of systems: let
$$ H = \sum_{i=1}^n H_i (x_i, p_i) \ \ \ , \eqno(3.1) $$ $x_i$, $p_i$
denoting positions and momenta of $n$ clusters of Schr\"odinger
particles. Let
$$H_i = H^0_i + V_i(x_i) $$ with $H^0_i$ the kinetic energy operator
of the $i$ cluster and $V_i(x_i)$ its interaction potential, which we
assume small with respect to $H^0_i$ in the sense of bilinear forms
[24], so that $H_i$ and $H$ are uniquely defined as self--adjoint
operators.  All $H_i$ are then bounded below and their point spectrum is
assumed to include non degenerate eigenvalues $\lambda_i^k$,
with eigenvectors $\psi_i^k$.  Any eigenstate of $H$, with
eigenvalue $E$ is then of the form
$$ \psi_E = \sum_{s=1}^N c_s \prod_{i=1}^n \psi_i^{k(i,s)} \eqno(3.2)
$$ with
$$ \sum_{i=1}^n \lambda_i^{k(i,s)} = E  \ \ \ \ \ \forall s = 1 \ldots N
\eqno(3.3) $$
The QM correlation functions of position observables on
such states are immediately calculated as
$$ (\psi_E, f_1(x_1(t_1)) \cdots f_n(x_n(t_n)) \psi_E) = $$
$$= (\psi_E,
e^{iH_1 t_1} \, f_1(x_1) \, e^{-i H_1 t_1} \cdots e^{iH_n t_n} \, f_n(x_n)
\, e^{-i H_n t_n} \psi_E) = $$
$$ = \sum_{s,p} {\overline c_s} c_p \prod_{i=1}^n (\psi_i^{k(i,s)} ,
f_i(x_i) \, \psi_i^{k(i,p)}) \; e^{i (\lambda_i^{k(i,s)} -
\lambda_i^{k(i,p)})t_i} \eqno(3.4) $$
and are therefore finite
combinations of products of trigonometric functions of $t_1 \ldots t_n$. In
particular, for $n=2$, and $c_s$ real, they are of the form
$$ (\psi_E, f_1(x_1(t_1)) f_2(x_2(t_2)) \psi_E) = \sum_{s=1}^N C_s
\cos \omega_s (t_1 - t_2) \eqno(3.5) $$ with $C_s$ real.

\msk
The derivation of correlation functions for position variables at
different times is almost trivial for Bohm Mechanics, as a consequence
of the stationarity assumption.  For $N$ particles, Bohm mechanics is in
fact defined by a probability distribution at $t=0$
$$\rho (x_1, \ldots , x_N) = |\psi(x_1 , \ldots , x_N)|^2  \eqno(3.6) $$
and by evolution equations
$$ {dx_i (t) \over dt} = {1 \over m_i} \; Im { \partial \over
\partial x_i } \log \psi( x_1 , \ldots , x_N , t) \ \  ,  \eqno (3.7) $$
$\psi (x_1 , \ldots , x_N , t) $ denoting a solution of the Schr\"odinger
equation
$$ i {\partial \over \partial t} \psi = \sum_i - {1 \over 2 m_i }
 \;  \Delta_i  \psi + V(x_1, \ldots  , x_N) \psi \eqno(3.8) $$
In the case of a stationary state, since $\psi$ can be taken to be real,
the velocity field on the r.h.s. of eq.(3.1) vanishes identically
and therefore the trajectories $x_i(t)$ are simply constant.
The probability distribution for position observables
$x_1 (t_1)$, $\ldots$, $x_k (t_k)$ {\it is therefore the same} as
the distribution for $ x_1 (0)$, $\ldots$, $x_k (0)$.
From eq.(3.4) we have therefore

\bsk\ni
{\bf Proposition 3.1}
{\it The Bohm probability distribution for
position observables of non interacting subsystems at different
times, for a stationary state $\psi$ of a system with Hamiltonian
of the form (3.1)
coincide with the quantum mechanical result, eq.(3.4)
if and only if $\psi$ is a product},
$$\psi (\xn) = \psi_1^{k_1} (x_1) \cdot \ldots
\cdot \psi_n^{k_n} (x_n)   \eqno(3.9)   $$

\msk\ni
E.g., for a system of two independent harmonic oscillators, with
Hamiltonian
$$ H  \equiv   { p_1^2 \over 2 } +  {\omega^2 x_1^2 \over 2} +
  { p_2^2 \over 2 } +  { \omega^2 x_2^2 \over 2}  $$
in the stationary state
$$  \psi \equiv {1 \over \sqrt 2} (\psi_0 (x_1) \psi_1 (x_2)  +
\psi_1 (x_1) \psi_0 (x_2)) \ \ \ , $$
$\psi_0$ denoting the ground state and $\psi_1$ the first excited
state of $H_0$, the Bohm probability distribution for
$x_1(t) , x_2(s)$ is independent of times;
the quantum mechanical expectation of their product is however
$$ (\psi \, , \, x_1(t) \, x_2(s) \, \psi) =
{ \cos \omega (t-s) \over 2 \omega } \ \ \ \  . $$

\bsk
In the case of Nelson Mechanics, probability
distributions for positions at different times are obtained from
solutions of stochastic equations.
For a system of $N$ particles, with positions $x_i(t)$,
Nelson Mechanics is defined [5] by the forward stochastic equations
$$ d x_i = b_i \; dt + dw_i  \eqno(3.10)  $$
with $dw_i$ independent white noises,
$$ b_i(\xN,t) = {1 \over m_i} \; (Re + Im) \ddxi \log \psi(\xN,t)
\eqno(3.11) $$
and $\psi$ a solutions of the Schr\"odinger equation, eq.(3.8).
As in Bohm Mechanics, the probability density is assumed to coincide
with $|\psi|^2$ at some time, and therefore at all times, as a
consequence of eqs. (3.10), (3.11) [5].

\msk
For non interacting clusters of particles, i.e. for $ V(\xn) = \sum_i
V_i(x_i) $, $x_i$ denoting the position vector of a cluster of
particles, if $\psi(\xn,0)$ is a product of wave functions, the same
holds for all times,
$$ \psi(\xn,t) = \prod_i \psi_i (x_i,t) \eqno(3.12) $$
It follows that the drift for each cluster of particle only depends on
position
variables within the cluster,
$$ b_i = b_i(x_i) \eqno(3.13) $$
so that equations (3.10) split into subsets.
Each of them gives rise to the Schr\"odinger evolution for the
particle density in the corresponding cluster and therefore the solution of
eqs.(3.10) (see below for its existence and uniqueness) gives a joint
probability density for $\xntn$
$$ \rho(\xntn) =
    \prod_i |\psi_i (x_i,t)|^2 \eqno(3.14) $$
It follows that for product wave functions the predictions of Nelson
Mechanics coincide with those of Quantum Mechanics, also for positions
at different times.

\msk
The situation is completely different for wave functions which are not
a product. In this case, even for stationary states, eqs.(3.10) are
not independent and their solution must be examined more closely.
Let us consider a (normalized) stationary state
$\psi_E(\xN)$ of a Hamiltonian of the form (3.1).
The Nelson forward stochastic equations are
$$ d x_i = {1 \over m_i} \; \ddxi \log \psi_E (\xN) \; dt + dw_i
\eqno(3.15)$$
and coincide, by reality of $\psi$, with the backward
equations.

If $\psi_E$ is not a ground state, it vanishes on surfaces of
codimension $1$, so that $\psi_E^{-1} \ddxi \psi_E$ is singular on such
surfaces.  Existence and uniqueness of the solution of eqs. (3.15),
i.e.  of the stochastic processes associated to eqs. (3.15) does not
therefore follow from standard results
\citaref{E. B. Dynkin, {\it Markov Processes}, Springer, 1965}.
A (unique) solution has been constructed by Carlen
\citaref{E. A. Carlen, Conservative diffusion, {\it Commun.
Math. Phys.} {\bf 94}, 293 (1984)}\
for a large class of Schr\"odinger
wave functions $\psi_E (x_1 \ldots x_N , t)$, by approximating singular
drifts with regular ones.

Carlen's construction is very general and solves Nelson's equations
also in the case of time dependent wave functions.  In the case of
stationary states, the solution has been discussed in
refs. \citaref{M. R\"ockner, T. S. Zhang, Uniqueness of Generalized
Schr\"odinger Operators and Applications, {\it Journ. Funct. Analysis} {\bf 105}, 187
(1992); {\bf 119}, 455 (1994)}
\citaref{A. Eberle, {\it Uniqueness and non--uniqueness
of semigroups generated by singular diffusion operators},
Springer, Lecture Notes in Math. 1718 (1999)}\
and shown to be given by the unique
Markov process with generator extending the Fokker--Planck operator
[20] associated to eqs. (3.15).
Such a Markov process is well defined, and actually generated
[22] by the Friedrichs extension
\citaref{T. Kato, {\it Perturbation theory for linear operators}, Springer,
1980}\
of the Fokker--Planck operator for all $\psi(\xN)$ in $L^2 (\reali^{N})$ with
$L^2$
derivatives.

It is easy to show that, as a consequence of the general structure of the
semigroup associated to a Markov process, invariant under time
reflection as required by coincidence of forward and backward drifts,
with stationary measure $\dmps$, the corresponding predictions for
correlations of non interacting subsystems at different times are not
compatible with the Quantum Mechanical result, eq. (3.4), unless
$\psi(\xn)$ is a product, eq. (3.9).  For completeness, in the
Appendix the general form of correlations at different times will be
given in terms of QM Hamiltonians; moreover, the Carlen construction
will be explicitly performed for a system of two harmonic oscillators,
resulting in disagreement with the QM predictions.

Let us therefore consider the structure of the unique solution
of eqs. (3.15). It is given [20] [23] by  a positivity
preserving semigroup $A(t)$, $ t \in [0,
\infty)$ acting in $L^2 (\reali^N , |\psi_E|^2 dx_1 \ldots dx_N)$ such
that
$$\int d\mu \; f(x(t)) \; g(x) = \int f \;  A(t) g \; \dmps
\ \ \ \ \ \ \  .  \eqno(3.16) $$
More generally, for $t_k > t_{k-1} >  \ldots  > t_1 $,
$$\int d\mu \; f_1 (x(t_1)) \ldots  f_k(x(t_k)) = $$
$$ = \int f_k (x) \;  A(t_k - t_{k-1}) (f_{k-1}(x)
 \ldots A(t_2 - t_1) f_1 (x)) \ldots )  \; \dmps  \ \  .
\eqno(3.17) $$
Stationarity of $\dmps$ and time reflection symmetry
imply $A(t) 1 = 1$ and hermiticity of $A(t)$, so that
$$ A(t) = \exp{-tT} \eqno(3.18)$$
with $T$ selfadjoint. From the spectral representation of the r.h.s.
of eq.(3.18) it follows that $T$ is non negative;
in fact, positivity of $A(t)$ implies boundedness in
$t$ of $||A(t)f|| $ for $f$ in a dense subspace since, for bounded
$f$,
$$ || A(t) f ||^2_{L^2 (\dmps)} = (f, A(2t)f) \leq (|f|, A(2t) |f|) \leq $$
$$ \leq \sup_x |f| \; (1, A(2t) |f|) \leq \sup_x |f| \;  ||f||  \ \ . $$
Restricting the attention to position variables of different non
interacting clusters of particles we have therefore

\bsk\ni
{\bf Proposition 3.2} {\it For any stationary state
$\psi_E$ of a system with Hamiltonian of the form (3.1),
the solution of the Nelson equations (3.15)
defines probability distributions, for position observables of non
interacting subsystems at different times,
of the form (3.17), (3.18).
In particular, the two--particle correlations are given by
$$\int d\mu \; f (x_1 (t)) \; g(x_2)) =
   \int  f(x_1) \;  \exp{-tT} \,  g(x_2)  \ |\psi_E(x_1,x_2)|^2
\, dx_1 \, dx_2  = $$
$$ = \int_0^\infty  \exp{-\lambda t} \; d\nu_{f,g} (\lambda)
\ \ \ \   ,  \eqno(3.19)$$
with $d\nu_{f,g}$ a complex measure. They coincide with the
QM result, eq. (3.4), if and only if $\psi$ is a product}.

\msk\ni
{\bf Proof}.
The spectral representation of $T$ gives rise to the complex measure
$$ (f, dE_T(\lambda) g) = d\nu_{f,g}(\lambda)$$ and to the
representation (3.19).
If a function $F(t)$ is of the form (3.19), then
$$ \lim_{t \to \infty} F(t) = \nu( \{ 0 \} ) \ \ \  ; $$
if $F(t)$ is also of the form (3.4), it must be a constant and
therefore $\psi_E$ is a product, as in Proposition 3.1.

\goodbreak
\bsk\bsk\ni
{\bf 4. No classical theory can reproduce QM correlation functions of
compatible observables at different times}.

\bsk\ni
We will show that the above discrepancy between Stochastic Mechanics
and QM is not special to Bohm's and Nelson's theories, by proving that
no classical probability model can reproduce the QM correlation
functions of compatible position observables at different times for
suitable, even very elementary, systems.

Consider a system composed of two identical sybsystems, with
Hamiltonian
$$ H = K(x_1, p_1) + K(x_2, p_2) \equiv K_1 + K_2  \eqno(4.1) $$
and let $E_0$, $E_1$ be two eigenvalues of $K$, with
eigenvectors $\psi_0$, $\psi_1$.
Let $f(x)$ be a measurable function, taking values $1$ and $-1$,  and
consider the correlation functions
$$ (\psi, f(x_1,t) \; f(x_2,s) \psi)  \eqno(4.2) $$
with
$$\psi = 1/\sqrt{2} \; (\psi_0(x_1) \psi_1(x_2) -
\psi_1(x_1) \psi_0(x_2))  \eqno(4.3)$$
and $f(x_i,t)$ the QM observable $f(x_i)$ at time $t$:
$$ f(x_i,t) = e^{itH} \, f(x_i) \, e^{-itH} = e^{itK_i} \, f(x_i) \,
e^{-itK_i} \eqno(4.3) $$

All the observables $f(x_i,t)$, $i = 1,2$, $t \in \reali$, take values
in $\{ -1,1 \}$. Given $N$ such observables, the most general
classical probabilistic theory consists therefore of a probability
measure on the space $\{ -1 ,1 \}^N $.  Following the discussion of
Bell inequalities in ref. \citaref{A. S. Wightman, Some comments on
the quantum theory of measurement, in {\it Probabilistic methods in
Mathematical Physics, Siena, 6--11 May 1991}, F. Guerra,
M. I. Loffredo, C. Marchioro Eds., World Scientific, 1992}, we will
prove the following:

\msk\ni
{\bf Proposition 4.1} {\it There exist times $t_i$, $s_i$, $i=1,2$
and QM models such that the set of QM predictions for the correlation
functions
$$ (\psi, f(x_1,t_i) \; f(x_2,s_j) \psi)  $$
cannot be reproduced by any probability assignment, i.e. there
does not exist a non negative probability distribution $p$, on variables
$\sigma_k$ and $\tau_l$ taking values in $\{ -1,1 \} $,
$$ \sum_{\sigma_k = \pm 1 \ \tau_l = \pm 1 }
p(\sigma_1, \sigma_2, \tau_1, \tau_2) = 1  \eqno(4.4)$$
such that}
$$ (\psi, f(x_1,t_i) \; f(x_2,s_j) \psi)  =
\sum_{\sigma_k = \pm 1 \ \tau_l = \pm 1 }
p(\sigma_1, \sigma_2, \tau_1, \tau_2) \  \sigma_i \, \tau_j
\  \equiv  \  <\sigma_i \, \tau_j>   \ \ \  .    \eqno(4.5)$$

\msk\ni
{\bf Proof}.
Following [25], it is immediate to check that the function
$$ \sigma_1 \tau_1 + \sigma_2 \tau_2 + \sigma_2 \tau_1 - \sigma_1 \tau_2 $$
takes values in $ [-2,2]$, so that, for any probability distribution,
$$ |<\sigma_1 \tau_1 + \sigma_2 \tau_2 + \sigma_2 \tau_1 - \sigma_1 \tau_2>|
\leq 2    \eqno(4.6)$$
Eq. (4.5) cannot therefore hold for QM models such that
$$ (\psi, f(x_1,t_1) \; f(x_2,s_1) \psi) +  (\psi, f(x_1,t_2) \; f(x_2,s_2)
\psi)  + $$
$$ + (\psi, f(x_1,t_2) \; f(x_2,s_1) \psi) -  (\psi, f(x_1,t_1) \;
f(x_2,s_2) \psi) < - 2
\eqno(4.7) $$
In order to construct such models, let
$$ f(x) \equiv sign(x) \ \ \ \  ,    $$
$K$ a selfadjoint operator with
$$ K \psi_i = E_i \psi_i \ \ \  ,  \ \ \ \  i = 0,1  \eqno(4.8) $$
and $\psi_i (x)$ real functions satisfying
$$  \psi_0(x) = \psi_0 (-x) \ \ \ , \ \ \ \ \psi_1(x) = - \psi_1 (-x) \ \ \
, \ \ \ \
   \int dx \; |\psi_i(x)|^2 =1  $$
It then follows
$$ (\psi_0, f(x) \psi_0) = (\psi_1, f(x) \psi_1) = 0   \eqno(4.9) $$
$$ (\psi_0, f(x) \psi_1) \equiv  \alpha \ \ \ , \ \ \ \ -1 \leq \alpha \leq
1  \eqno(4.10)$$
Let $ \omega \equiv E_1 - E_0 $.
By eqs (4.3), (4.8), (4.9), (4.10), the above expectation values are given
by
$$ (\psi, f(x_1,t) \, f(x_2,s) \psi)  =
   (\psi \, , e^{iK_1t} \, e^{iK_2s} \, f(x_1) \, f(x_2) \, e^{-iK_1t} \,
e^{-iK_2s} \, \psi)  = $$
$$ = -  Re\,  (\psi_0(x_1) \psi_1(x_2) \,  ,
      \, e^{iK_1t}  \, e^{iK_2s} \, f(x_1) \,  f(x_2)  \, e^{-iK_1t}  \,
e^{-iK_2s} \,
      \psi_1(x_1) \psi_0(x_2)) = $$
$$ = -  \alpha^2 \cos{ \omega (t-s)} $$
Choosing $t_1 =0$, $t_2 = \pi/2 \omega$,
$s_1 = \pi/4 \omega$, $s_2 = 3 \pi / 4 \omega$, the
result for the l.h.s. of eq.(4.7) is
$$ - 4 \, \cos (\pi/4) \; \alpha^2 = - 2 \sqrt{2} \; \alpha^2  $$
and satisfies inequality (4.7) if
$$\alpha^2 > \sqrt{2} / 2 \eqno(4.11) $$

The construction of models satisfying inequality (4.11) is
straightforward; e.g., for a free particle on an interval $[-L,L]$,
with Dirichlet boundary conditions, taking $\psi_0$ and $\psi_1$ as
the lowest energy states, one obtains $\alpha^2 = 8/(3 \pi) > \sqrt{2}
/ 2$. By taking the lowest energy states of double well potentials one
can obtain $\alpha = 1 -
\varepsilon$, with $\varepsilon > 0$ arbitrarily small. $\alpha = 1$
would correspond to the maximal violation of the Bell inequalities
(4.6) by QM states [25].

\msk
The above derivation strictly reproduces the discussion of Bell
inequalities for spin variables, $f(x_i, t)$ playing the r\^ole of
spin variables and eq. (4.3) corresponding to a
state with zero total angular momentum. It shows that the
existence of a classical probabilistic representation has nothing to
do with the choice of position variables and rather depends, as
discussed in Proposition 2.6, on the exclusion of observations at
different times, giving rise to a transitive compatibility relation.

\vfill\eject

\centerline{\bf Appendix }

\bsk
\ni
Explicit formulas will be given here for the correlation functions at
different times on stationary states in Nelson Stochastic Mechanics,
based on the general results of [22].  Moreover, the Carlen
construction will be performed directly, by elementary methods, for a
system of two harmonic oscillators and the result will be compared
with the corresponding QM predictions.

\bsk\ni
{\bf A1}. Let us consider a Hamiltonian
$$ H = - \sum_{i=1}^N { \Delta_i \over 2 m_i } \, + V(x_1 \ldots x_N)
\eqno(A.1) $$
with $V = V_1 + V_2$, $V_1$ small in the sense of forms
with respect to the kinetic term, $V_2 \in C^\infty$ and bounded
below, and $\psi(\xN)$ a regular ($C^2$) eigenfunction of $H$.
The assumptions on $V$ imply that $\psi$ is in the domain of the form
defined by the Laplacean, so that it has $L^2$ derivatives.  It
follows [22] that the solution of eqs. (3.15) is
given by the semigroup generated by the Friedrichs extension [24] of
the Fokker--Planck operator
$$ F \equiv  \sum_{i=1}^N {\Delta_i \over 2 m_i} \,  + \sum_{i=1}^N
{\partial \over \partial x_i} \, \log
\psi (x) \, {\partial \over \partial x_i}  \ \ \ \ \  ,  \eqno(A.2)  $$
defined in $L^2(\reali^{3N}, \, \psi^2 \, dx_1 \ldots dx_N)$ on the
domain ${\cal D}_0$ of $C^2$ functions with compact support,
vanishing in a neighbourhood of the set of zeros of $\psi$.
A more explicit representation can be obtained by using the isometry
$$ W : L^2 (\reali^{3N}, \, \psi^2 \, dx_1 \ldots dx_N)
\longrightarrow L^2(\reali^{3N} , \, dx_1 \ldots dx_N)    $$
defined by
$$ W \, f(x) \equiv |\psi(x)| \, f(x)  \ \ \ \ \  .    \eqno(A.3)$$
In fact, for every $f$ in  ${\cal D}_0$,
$$  - W^{-1} \, (- \sum_{i=1}^N { \Delta_i \over 2 m_i} \, + V(x)) \,
  W \, f(x) =
 |\psi|^{-1} \sum_i ( {\Delta_i \over 2 m_i} \, |\psi| ) \; f
+ \sum_i {1 \over 2 m_i} \Delta_i \, f + $$
$$ + |\psi|^{-1} \sum_i {\partial |\psi| \over \partial x_i} {\partial f
\over
\partial x_i } - V f = (F - E) \, f(x) $$
where we have used the eigenvalue equation $ H \psi = E \psi $, so that,
on ${\cal D}_0$,
$$ F = - W^{-1} \, (H-E) \, W  \eqno(A.4) $$
Selfadjoint extensions of $F$ in
$L^2(\reali^{3N},\, \psi^2 \, dx_1 \ldots dx_N)$ correspond
therefore to selfadjoint extensions of $H$ from the domain
$ W {\cal D}_0$:

\bsk\ni
{\bf  Proposition A.1} {\it Every selfadjoint extension $\tilde{F}$ in
$L^2(\reali^{3N},\, \psi^2 \, dx_1 \ldots dx_N)$
of the operator $F$ is of the form
$$ \tilde{F} = -W^{-1} \, (\tilde{H} - E) \, W \eqno(A.5)$$
with $\tilde{H}$
a self-adjoint extension of $H$, defined on $ W {\cal D}_0$ in
$L^2(\reali^{3N}, \, dx_1 \ldots dx_N)$, $W$ the isometry given by
eq. (A.3).  The Friedrichs extension of $F$ is obtained, through
eq. (A.5), from the Friedrichs extension $H^F$ of $H$.
$H^F$ is essentially selfadjoint on the domain of $C^\infty $
functions with compact support vanishing on the zeros of $\psi$.
The semigroup $A(t)$, eq. (3.18) can therefore be written
$$ A(t) = W^{-1} \, e^{- t \big( -
(\sum_i \Delta_i / 2m_i)_0 + V - E \big) } \, W  \ \ \ \  ,  \eqno(A.6) $$
$(\sum_i \Delta_i / 2m_i)_0$ denoting the quantum mechanical kinetic energy
operator on $C^\infty$ functions with compact support vanishing
on the zeros of $\psi$}.

\msk\ni
{\bf Proof}.  The isometry $W$ maps the deficiency subspaces of the
hermitean operator $F$, on ${\cal D}_0$, into those of $H$, on
$W {\cal D}_0$. $W$ also maps form domains into form domains.  By
regularity of $\psi$, the domain $W {\cal D}_0$ consists of
all $C^2$ functions with compact support not intersecting the zeros
of $\psi$. The Friedrichs extension of the kinetic term $H_0$ gives
therefore the sum of Laplaceans with Dirichlet boundary conditions and,
by an elementary computation of deficiency spaces, this operator
is essentially selfadjoint on the domain of $C^\infty $ functions
with compact support vanishing on the zeros of $\psi$.
Since $V_2$ is regular and bounded below, $H_0 + V_2$ remains essentially
selfadjoint on the same domain (see \citaref{M. Reed, B. Simon,
{\it Methods of Modern Mathematical Physics, Vol. II}, Academic Press, 1975},
 Theorems X.27, X.28), and coincides there
with its Friedrichs extension from $W {\cal D}_0$.
$V_1$ is form--small with respect to $H_0$ and therefore
with respect to $H_0 + V_2$, as a consequence of $V_2$
being bounded from below; $H_0 + V_1 + V_2$ is therefore essentially
selfadjoint on the same domain and coincides there with the Friedrichs
extension of $H$ from $W {\cal D}_0$.

\bsk\ni
{\bf A2}. We shall now rederive explicitly eq. (A.6) in a simple case. i.e
for
the first excited state of a harmonic oscillator, by performing the
Carlen construction by direct and elementary methods. The result will
turn out to be enough for the discussion of compatible position
observables of a pair of harmonic oscillators.

Let $H$ the Hamiltonian of a one dimensional harmonic oscillator,
$$ H = - {1 \over 2} \big( {\partial^2 \over \partial x^2} -
\omega^2 x^2 \big)    \ \ \ \ . \eqno(A.7) $$
The Nelson equations corresponding to the first excited state
$$ \psi_1(x) = \big( {4 {\omega}^{3} \over \pi} \big)^{1/4} x \ e^{-
{\omega \over 2} x^2} \ \ \ , \ \ \ \ H \psi_1 = E_1 \psi_1 \ \ \ ,
\eqno(A.8) $$
define a Fokker--Planck operator $F$ of the form
$$ F = {1 \over 2} {\partial^2 \over \partial x^2} +
{\partial \over \partial x} \, \log \psi_1 (x) \,
{\partial \over \partial x} \ \ \ \  .  \eqno(A.9) $$
The drift term in eq.(A.9) is singular at the origin and a possible
domain ${\cal D} \subset L^2(\reali, \psi_1^2 dx)$ of $F$
consists of (smooth) functions vanishing at the origin. It is
easy to see that it is a domain of hermiticity but not of
(essential) selfadjointness for $F$.

We will perform explicitly Carlen's construction [21],
which give the sto\-cha\-stic processes associated to singular
drifts in terms of a sequence of regular approximants.
The result will be given, as in Proposition (A.1), by the semigroup
generated by the Hamiltonian operator (A.7),
with Dirichlet boundary conditions at $0$, the node of $\psi_1$.
Following Carlen, we introduce strictly positive regular functions
$\psi_0^{\eps}(x) $ approximating $\psi_1$ and discuss convergence of
the corresponding Nelson processes.
Let $\psi_0^{\eps}(x) \equiv |\psi_1(x)|$, for $|x| > \epsilon$, and
$ \psi_0^{\eps}(x) \equiv g^\eps (x)$, for $|x| \leq
\epsilon$, $g^\eps (x)$ a positive even function in $C^2$.
$g^\eps$ and its derivative are assumed to coincide with $|\psi_1|$ at
$x = \pm \eps$, and to satisfy
$$ c_1 \, \eps^{-2} g^\eps (x) \leq d^2 / dx^2 \; g^\eps (x) \leq
c_2 \, \eps^{-2} \,  g^\eps (x)    \ \ \ \  .  \eqno(A.10)   $$
E.g., $g^\eps$ can be taken of the form
$$  g^\eps(x)  = a^\eps \, \cosh (b^\eps x)    $$
with $ a^\eps = O(\eps)$, $b^\eps = O(1/\eps)$.
Introducing
$$ \delta V^\eps(x)\equiv  {\partial_x^2 g^\eps (x)  \over  2 g^\eps(x)}
- { \omega^2 x^2 \over 2} + E_1 \ \ \ \  , \ \ \ \ \  |x| \leq \eps$$
and zero otherwise, we have
$$ H^\eps  \psi_0^\eps = E_1 \psi_0^\eps  \ \ \ ,  \eqno(A.11)  $$
with
$$ H^{\eps} \equiv H + \delta V^{\eps}(x)    \eqno(A.12)$$
Being positive, $ \psi_0^\eps$ is the ground state of
$H^{\eps}$. The drifts
$$ b^{\eps}(x) = {\partial \over \partial x} \log  \psi_0^{\eps}(x)
\eqno(A.13) $$
are regular, so that the
Fokker--Planck operator $F^{\eps} = \Delta/2 + b^{\eps}(x)
\partial_x$ is essentially selfadjoint on the domain of
smooth functions with compact support in $L^2(\reali,(\psi_0^{\eps})^2
dx)$ and generates the semigroup $A^\eps (t)$ associated to
the Nelson stochastic processes with drift $b^{\eps}$.
As in Proposition A.1, we have the representation
$$ A^{\eps}(t) = {1 \over  \psi_0^{\eps}} \, e^{-(H^{\eps} - E_1) t}
\, \psi_0^{\eps} \eqno(A.14)  $$
Following Carlen's construction, we will discuss the convergence of
$A^\eps (t)$ for $\eps \to 0$.

\bsk\ni
{\bf Proposition A.2} {\it There are sequences $\eps_k$ such that the
operators $e^{-H^{\eps} t}$ converge in norm in $L^2(\reali, dx)$. The
limit is unique and given by $e^{- \tilde{H}  t}$,
with $\tilde{H}$ the Hamiltonian (A.7) with Dirichlet boundary
conditions at $x = 0$.
The corresponding operators $A^{\eps_k} (t)$ converge
strongly on a dense domain in  $L^2(\reali,\psi_1^2 dx) $ to }
$$  {1 \over | \psi_1 | } \, e^{- (\tilde{H} - E_1) t}\,
| \psi_1 |  =
  {1 \over  \psi_1  } \, e^{- (\tilde{H} - E_1) t}\,
 \psi_1   \eqno(A.15)  $$

\msk\ni
{\bf Proof}. By eq.(A.10), for $\eps$ small,
$$ 0 \leq \delta V^{\eps}(x) \leq  c_2 / x^2 $$
and therefore
$$ H  \leq H^{\eps} \leq H  + c_2 / x^2 \equiv H^{>}
\ \ \ \  .  \eqno(A.16) $$
The operator $H^{>}$ is hermitian on the domain
${\cal D}^{>} = \{ f \in C^{\infty}_0 , f(0) = f^{'}(0) = 0 \}$
and its Friedrichs extension is greater than $H^{\eps}$, so that, by
the minimax principle, the eigenvalues $E_n^\eps$, $n\geq 0$,
of the operators in eq. (A.16) satisfy
$$ E_n  \leq E_n^{\eps} \leq E_n^{>} $$
There are therefore sequences $\{ \eps^i_k \}$, $\{
\eps^i_k \} \subset \{ \eps^j_k \}$ if $i > j$, such that
$E_i^{\eps^i_k}$ converge as $k\to \infty$.
For $\{ \eps_k \} \equiv \{ \eps_k^k \} $,
we have convergence of all the eigenvalues for $k \to \infty$.
By Lemma A.3 below, the eigenfunctions $\psi_n^k$ of $H^{\eps_k} $
converge to the eigenfunctions $\tilde\psi_n$ of $\tilde{H}$. Moreover,
$$ e^{-H^{\eps_k}t} = \sum_{n=0}^N e^{- E_n^{\eps_k} t}
|\psi_n^k><\psi_n^k| + R_N^k  \eqno(A.17) $$
with $ \| R_N^k \| \leq e^{- E_{N+1} t} $ as a consequence of eq.(A.16).
The operators
$ e^{-H^{\eps_k}t} $ converge therefore in norm, for $k \to \infty$,
to $ e^{-\tilde{H} t}$.
The ground state wave functions $ \psi_0^k$ are positive and, by Lemma A.3,
converge to $ |\psi_1|  $ uniformly. The operators
$$ A^{\eps_k} =  {1 \over  \psi_0^k } \; e^{- H^{\eps_k} t}\; \psi_0^k =
{1 \over |\psi_1|} \; \big ( {|\psi_1| \over  \psi_0^k } \; e^{- H^{\eps_k}
t} \;
 {\psi_0^k \over  |\psi_1| } \big ) \; |\psi_1| \eqno(A.18)  $$
therefore converge, on the dense domain of regular functions with compact
support excluding  $0$ in $L^2(\reali,\psi_1^2 dx) $, to
$$  {1 \over | \psi_1 | } \; e^{- \tilde{H} t}\; | \psi_1 | =
  {1 \over  \psi_1  } \; e^{- \tilde{H} t}\;  \psi_1  \ \ \ \ \  ,    $$
the last equation following form the fact that $L^2 ((0, \infty))$ and
$L^2 ((- \infty, 0))$ are left invariant by $e^{- \tilde{H} t}$.

\bsk\ni
{\bf Lemma A3} {\it If a sequence of eigenvalues $E_n^{\eps_j}$ of the
Hamiltonians defined by eq.(A.12), $n$ fixed, converges for $j \to \infty$,
then the corresponding normalized eigenfunctions $\psi_n^{\eps_j}$
converge to the eigenfunctions of $\tilde{\psi}_n $ of $\tilde{H}$}.

\msk\ni
{\bf Proof}. Since $V^\eps$ is even, the eigenfunctions  $\psi^j \equiv
\psi_n^{\eps_j}$
($n$ fixed) have definite parity, so that the analysis can be restricted to
$x \in [0,\infty)$ and we can assume
$ \psi^j(x) > 0 $ for $ x > 0$ sufficiently small.
For $\eps_j$ small and for $ x$ small in $(0,\eps_j) $, the
Schr\"odinger equation gives
$$ {1 \over 2}  \;{ d^2 \over dx^2 } \;
\psi^j (x) =  ( {d^2 g^\eps (x)  \over dx^2 } (2 g^\eps(x))^{-1}
- E_n^{\eps_j} ) \; \psi^j (x) \geq \; {c \over 2 \, \eps_j^2} \; \psi^j (x)
\eqno(A.19)$$
for $c < c_1$ and $\eps_j$ small.
$d/dx \, \psi^j (x)$ is therefore positive for $x >0$ small, Eq. (A.19)
extends to all $x$ in $(0, \eps_j)$, so that $d/dx \,  \psi^j (x)$ is
increasing in
$(0,\epsilon)$ and we can assume the normalization
$$  d/dx \, \psi^j (\eps_j) = 1  \ \ \ \  .  \eqno(A.20)$$
Moreover, eq.(A.19) implies
$$ d/dx \; \psi^j(x) \geq {c \over \eps_j^2} \, x \, \psi^j(0)  $$
so that
$$ \psi^j(0) \, \leq \, \eps_j / c \ \  d/dx \, \psi^j(\eps_j) =  \eps_j / c
\eqno(A.21) $$
and therefore
$$ 0 \leq \psi^j (\eps_j) \leq (c^{-1} +1) \eps_j \eqno(A.22)$$
For $x > \eps_j$, $ \psi^j(x)$ satisfies the equation
$$ - 1/2 \; d^2/dx^2 \, \psi^j (x) + \omega^2 /2 \; x^2 \, \psi^j (x) =
E_n^{\eps_j} \, \psi^j (x) \ \ \ \  ,   \eqno(A.23)$$
with the boundary conditions (A.20), (A.22) at $x=\eps_j$.
Eqs. (A.20), (A.22), (A.23) imply pointwise convergence of
$\psi^j(x + \eps_j)$,  $x \in [0, \infty)$, uniform on compact sets.
Since for $x$ large we have the bound
$$ |\psi^j (x)| \ < \ a \ e^{-bx} \ \ \ \  ,  \eqno(A.24)$$
this implies convergence of $\psi^j(x)$ in $L^2 ([0, \infty)$, to
a limit $\psi_n$ satisfying eq.(A.23) with $E_n \equiv \lim_j E_n^{\eps^j}$
and $\psi_n(0)=0$. By counting the number of zeros, it follows that all the
eigenfunctions of $\tilde H$, the Hamiltonian with Dirichlet boundary
conditions in $x =0$, are obtained as limits of $\psi_n^j$.
In order to prove eq.(A.24), since the wave functions $\psi^j (x)$
are real and have a finite number of zeros, we can assume $ \psi^j (x)> 0 $
for all $j$, for $x$ sufficiently large;
the Schr\"odinger equation (A.23) then implies, for $x > L$, $L$
sufficiently large
$$ d^2/dx^2 \, \psi^j (x) \ \geq \ M \; \psi^j (x) \; > \; 0  \eqno(A.25)
$$
with $M = \inf_{j, \, x>L} (\omega x^2- 2 E_n^{\eps^j}) $.
Eq.(A.25) implies  $d/dx \; \psi^j(x) < 0$ for $x>L$ and, by multiplication
with $d/dx \; \psi^j (x)$,
$$ d/dx \, (\psi^j)^2 \; \leq \; M \; (\psi^j)^2  $$
so that
$$ \log \, \psi^j (x) < \log \, \psi^j(L)  - M/2  \ x $$
and the result follows from convergence of $\psi^j(L)$ for $j \to \infty$.

\bsk\ni
{\bf A3}.
We will consider now a system of two independent one dimensional harmonic
oscillator in the state
$$  \psi =  {1 \over \sqrt 2} (\psi_0 (x_1) \psi_1 (x_2)  + \psi_1 (x_1)
\psi_0 (x_2)) \ \ \ \  ,   $$
$\psi_0$ and $\psi_1$ denoting respectively the ground state and the first
excited state of the operator $1/2 \, (p^2 + \omega^2 x^2)$.
With the change of coordinates
$$ x \equiv { x_1 + x_2 \over \sqrt{2}} \ \ \ \ \ \ \
y \equiv  { x_1 - x_2 \over \sqrt{2}}  \eqno(A.26) $$
the Hamiltonian
$H = {1 \over 2} (p_1^2 + p_2^2 + \omega x_1^2 + \omega
x_2^2)$  becomes
$$ H = - {1 \over 2} {\partial^2 \over \partial x^2} - {1 \over 2}
{\partial^2 \over \partial y^2} + {1 \over 2} \omega^2 (x^2 + y^2) \equiv
H_x + H_y $$
and the above state
$$ \psi(x,y) = {\omega \over \sqrt{\pi}} \ x \ e^{-{\omega \over 2}(x^2 +
y^2)} = \psi_1(x)\psi_0(y)  \eqno(A.27) $$

Since, in the new variables, the wave function is a product,
we can solve separately the stochastic processes associated to
$\psi_1(x)$ and $\psi_0(y)$ respectively. The $y$ subsystem is in the
ground state and therefore its evolution semigroup is
$A_y(t) = \psi_0(y)^{-1} \; e^{-(H_y - \omega/2) t} \; \psi_0(y)$.
On the other hand, the $H_x$ subsystem is in the first excited
state, so that Proposition A.2 applies and the associated semigroup
can be written
$$ A_x(t) = {1 \over \psi_1(x)} \  e^{- (\tilde{H}_x - E_1) t}
\  \psi_1(x) $$
with $\tilde{H}_x$ the Hamiltonian with Dirichlet boundary condition at
$x=0$. The semigroup associated to the stochastic process of the whole
system is then
$$ A(t) = {1 \over \psi(x,y)} \ e^{- (\tilde{H}_x - E_1) t}
\ e^{- (H_y - \omega/2) t} \ \psi(x,y) \eqno(A.28) $$

The expectation of the position observable $x_1(t) x_2(0)$
in Nelson Stochastic Mechanics can therefore be calculated as
$$ ( \psi, \, x_1(t) \, x_2(0) \, \psi )_{NM} = {1 \over 2} \Big( \psi(x,y)
,
\, (x+y) \, e^{-(\tilde{H}_x - \tilde{E}_0) t} \, e^{-(H_y - \omega /2 )t}
\, (x-y) \, \psi(x,y) \Big) \eqno(A.29)$$
The only non trivial term in the r.h.s. of eq.(A.29) is
$$ \Big( \psi(x,y) , \, x \, e^{-( \tilde{H}_x - \tilde{E}_0) t}
\, e^{-(H_y - \omega/2) t} \, x \, \psi(x,y) \Big) =
\Big( \psi_1(x) , \, x \, e^{-(\tilde{H}_x - \tilde{E}_0) t} \, x \,
\psi_1(x) \Big) = $$
$$ = \sum_{n \ {\rm odd}} |c_n|^2 \, e^{-(n-1) \omega t} \ \ \ \  ,  $$
$c_n$ denoting the coefficients in the expansion of $x \psi_1(x)$ in terms
of the eigenfunctions of $\tilde{H}_x$, which can be written
$$ c_n = \int_{-\infty}^{\infty} \tilde{\psi}^{+}_n(x) \; x \; \psi_1(x) dx
= 2
\int_0^{\infty} x \; \psi_n(x) \; \psi_1(x) dx$$
where the eigenfunctions $\tilde{\psi}^{+}_n(x)$,
$n \in \naturali$ odd, with eigenvalues $\tilde{E}_n = (n+1/2) \omega$
are the even continuation in $(- \infty,
0)$ of the eigenfunctions $\psi_n (x)$ of the harmonic
oscillator in $x \in (0, \infty)$
(odd continuations also define eigenfunctions, with the same eigenvalues,
but they are irrelevant for the above expansion).
The other terms are easily calculated and the result is
$$ ( \psi, x_1(t)x_2(0) \psi )_{NM} = {1 \over 2} \Big( - {e^{-\omega t}
\over 2 \omega} + \sum_{n \ {\rm odd}} |c_n|^2 e^{-(n-1) \omega t} \Big)
 \ \ \ \   .  \eqno(A.30) $$
The same expectation in QM is given by (see Section 3)
$$ ( \psi, x_1(t)x_2(0) \psi )_{QM} = {\cos \omega t \over 2 \omega} \ \ \ \
.   $$

\bsk\bsk
\centerline{{\bf Acknowledgements}}

\msk\ni
M.C. is grateful to INFN for financial support. G.M. thanks Franco
Strocchi for many illuminating discussions.

\vfill\eject

\immediate\closeout\fileref
\par\vfill\eject
\null\msk
\centerline{\bf References}
\bsk
\input ref.tmp

\bye